# Fast suppression of superconductivity with Fe site Ni substitution in Fe$_{1-x}$Ni$_x$Se$_{0.5}$Te$_{0.5}$ (x=0.0, 0.01, 0.03, 0.05, 0.07, 0.10 and 0.20) single crystals

P.K. Maheshwari[1,2], Bhasker Gahtori[1] and V.P.S. Awana[1,#]

[1]CSIR-National Physical Laboratory, Dr. K. S. Krishnan Marg, New Delhi-110012, India

[2]Academy of Scientific and Innovative Research, NPL, New Delhi-110012, India

**Abstract**

We report the effect of Ni doping on superconductivity of FeSe$_{0.5}$Te$_{0.5}$. The single crystal samples of series Fe$_{1-x}$Ni$_x$Se$_{0.5}$Te$_{0.5}$ (x=0.0, 0.01, 0.03, 0.05, 0.07, 0.10 and 0.20) are synthesized via vacuum shield solid state reaction route and high temperature heating followed by slow cooling. All the crystals of Fe$_{1-x}$Ni$_x$Se$_{0.5}$Te$_{0.5}$ series with x up to 0.20, i.e., 20% substitution of Ni at Fe site are crystallized in single phase tetragonal structure with space group *P*4/*nmm*. The electrical resistivity measurements revealed that $T_c$ decreases fast with increase of Ni concentration in Fe$_{1-x}$Ni$_x$Se$_{0.5}$Te$_{0.5}$. Namely the superconducting transition temperature ($T_c$) being defined as resistivity ($\rho$)=0 decrease from 12K to around 4K and 2K for x=0.01 and 0.03 samples respectively. For x=0.05 (5at% Ni at Fe site) though $T_c^{onset}$ is observed in resistivity ($\rho$-T) measurements but $\rho$=0 is not seen down to 2K. For x ≥ 0.07, neither the $T_c^{onset}$ nor $T_c^{\rho=0}$ is seen down 2K in $\rho$-T measurements. It is demonstrated that Ni doping at Fe site in FeSe$_{0.5}$Te$_{0.5}$ superconductor suppresses superconductivity fast. The rate of $T_c$ depression is albeit non monotonic. Summarily, a systematic study on suppression of superconductivity with Fe site Ni doping in flux free gown FeSe$_{0.5}$Te$_{0.5}$ single crystals is presented in the current communication.

[*]**Corresponding Author**
Dr. V. P. S. Awana: E-mail: awana@mail.nplindia.org
Ph. +91-11-45609357, Fax-+91-11-45609310
Homepage awanavps.webs.com

**Introduction**

Discovery of superconductivity in Fe based compounds, first in pnictides [1,2] and soon later in chalcogenides [3,4] had been a biggest surprise to condensed matter physics community in over two decades after the infamous high $T_c$ (*HTSc*) cuprates [5]. It appeared as if thunder struck again after 1986 [5] in year 2008 [2]. In fact curious commentaries were written to raise the national passions [6]. Well the Fe based compounds superconductivity is so very much important because the same being outside the conventional popular superconductivity theory of BCS [7] is as mysterious as the one of Cuprates. Both the Cuprates and Fe pnictides superconductivity yet lacks a theoretical explanation and hence is a big challenge to the condensed matter theorist's community. As far as the Fe based compounds are concerned, the Fe chalcogenides are the simplest ones [3,4].

One of the ways to look at the phenomenon of superconductivity is to probe the suppression of the same with on site substitutions. As mentioned above, among the Fe based superconductors, the chalcogenides ones are simplest ones, i.e., like their structure is planer. In Fe chalcogenides series the highest superconducting transition temperature ($T_c$) yet obtained is for $FeSe_{0.5}Te_{0.5}$, which is around 15K at ambient pressure [8-11]. As far as the onsite substitutions are concerned the Fe can be tried to be partially replaced by say Ni, Co, Cr and Zn within same crystal structure. In this direction, some reports do exist in literature on partial substitution of Fe by Ni [12,13], Co [12,14], Cr [14,15] and Zn [15,16] in Fe chalcogenide superconductors. Interestingly enough most of the reported literature [12-15] on Fe site 3d metal (Ni/Co/Cr/Zn) doping in Fe chalcogenide superconductors is on bulk polycrystalline samples and that also for select few compositions. For better understanding, not only the numbers of samples need to be studied with different dopant concentrations but also preferably the single crystalline ones instead of the bulk poly crystals. The single crystal growth of Fe chalcogenide superconductors i.e., $FeSe_{0.5}Te_{0.5}$ (15K) and FeSe (6K) is rather complicated and is mainly possible only through added flux (KCl/NaCl) [17-20]. However, very recently the flux free large single crystals of both $FeSe_{0.5}Te_{0.5}$ (15K) and FeSe (6K) were reported in literature [21-23]. Following the protocol [23], we have grown single crystals of series $Fe_{1-x}Ni_xSe_{0.5}Te_{0.5}$. Very recently, the pure (superconducting) and Fe site 10at% Ni doped (non superconducting) $FeSe_{0.5}Te_{0.5}$ flux free crystals of ours were exposed to ARPES (Angle Resolved Photo Electron Spectroscopy) studies [24]. A decrease in the size of hole pockets near the zone center and an increase in the size of electron pockets near the zone corner was seen with Fe site Ni doping in $FeSe_{0.5}Te_{0.5}$, thus suggesting an effective electron doping [24]. Here we report the growth and basic superconductivity characterization of $Fe_{1-x}Ni_xSe_{0.5}Te_{0.5}$ (x=0.0, 0.01, 0.03, 0.05, 0.07, 0.10 and 0.20) flux free single crystals and compare the same with similar bulk polycrystalline samples or flux assisted crystals. It is found that Fe site Ni suppresses superconductivity fast in $FeSe_{0.5}Te_{0.5}$ superconductor. To our knowledge this is *first study on flux free $Fe_{1-x}Ni_xSe_{0.5}Te_{0.5}$ single crystals*, the earlier ones are either on poly-crystals [12-15] or flux assisted tiny crystals [16,25].

**Experimental Details**

The constituent elements i.e., Fe, Se, Te and Ni of better than 4N purity taken in stoichiometric ratio i.e., $Fe_{1-x}Ni_xSe_{0.5}Te_{0.5}$ (x=0.0, 0.01, 0.03, 0.05, 0.07, 0.10 and 0.20),are mixed in Ar gas filled glove box, heated to $1000^0C$ for 24 hours with an intermediate step of $450^0C$ for 4 hours. The protocol is same as applied in ref. 23. Finally the furnace is slow cooled ($10^0C$/hour) down to room temperature. The crystals thus obtained were big in size (2cmx2cmx1/2cm). It looked like as if the whole melt is grown in single crystalline form. The representative photograph of the crystals is shown in Fig. 1. The morphology of the $Fe_{1-x}Ni_xSe_{0.5}Te_{0.5}$ crystals is seen by scanning electron microscopy (SEM) images being taken on a ZEISS-EVO-10 electron microscope, and the Energy Dispersive X-ray spectroscopy (EDAX) is employed for elemental analysis.X-ray diffraction (*XRD*) is done at room temperature using CuK$_\alpha$ radiation of wavelength 1.5418Å for structural analysis on Rigaku diffractometer. To ascertain the superconducting transition temperature ($T_c$), resistance versus temperature measurements were performed by four probe technique on quantum design Physical Property measurements system (*PPMS*-14 Tesla) - down to 2K at various applied magnetic fields.

**Results and Discussion**

Although the SEM images were taken for most of the studied crystals, the one being representative taken for $Fe_{0.95}Ni_{0.05}Se_{0.5}Te_{0.5}$ on a micron level (80µm) is shown in Fig. 2(a). The layer by layer laminar growth of the crystal is clearly seen. Very similar growth morphology images were seen by us recently for pure $FeSe_{0.5}Te_{0.5}$ [23]. The compositional analysis taken on the selected area marked in Fig. 2(a) is being depicted in Fig. 2(b). The compositional analysis revealed slight deficiency of Se but not the excess Fe. The Ni concentration is also found to be close to the nominal value. As mentioned, the SEM and EDAX was done for all samples and only the representative Figs as Fig. 2(a) for SEM and Fig. 2(b) for EDAX are given for the $Fe_{1-x}Ni_xSe_{0.5}Te_{0.5}$ sample. Successive doping of Ni in $FeSe_{0.5}Te_{0.5}$ crystal was ascertained from EDAX for all samples and the obtained compositions were close to the nominal one, besides the deficiency of Se up to 5-7% in all crystals.

Fig. 3 depicts the room temperature single crystal XRD of the studied $Fe_{1-x}Ni_xSe_{0.5}Te_{0.5}$ compositions. As seen from Fig. 3, all the crystals are grown in (00l) plane with clear appearance of predominant [001], [002] and [003] planes in XRD. It is found that the [00l] planes shifts towards higher angle with increase in Ni content in $Fe_{1-x}Ni_xSe_{0.5}Te_{0.5}$. This is a clear indication that with Ni doping in $Fe_{1-x}Ni_xSe_{0.5}Te_{0.5}$, the *c* lattice parameter decreases monotonically. This is in agreement with earlier reports on Ni doped $FeSe_{0.5}Te_{0.5}$ poly [12-15] or flux assisted tiny crystals [16, 25]. To know the lattice parameters and co-ordinate positions etc., we carried out the powder XRD on all the studied crystals of $Fe_{1-x}Ni_xSe_{0.5}Te_{0.5}$. The individual crystals were first gently powdered and then subjected to powder XRD at room temperature. The results of powder XRD for crushed $Fe_{1-x}Ni_xSe_{0.5}Te_{0.5}$ are shown in Fig.4. All the studied samples are crystallized in tetragonal structure with space group *P*4/*nmm*, without traces of secondary or un-reacted phases in the matrix. Lattice parameters and co-ordinate positions are obtained by Reitveld fitting of the observed XRD patterns. Obtained lattice parameters and co-ordinate positions for $Fe_{1-x}Ni_xSe_{0.5}Te_{0.5}$ crystals are given in Table 1. Both the *a* and *c* lattice parameters and hence the volume are found to decrease with Ni substitution at Fe site in $FeSe_{0.5}Te_{0.5}$ single crystals. To clearly visualize the decrease in lattice parameters, the zoom part of powder XRD depicting [003] plane is shown in inset of Fig. 4. The higher angle shift of the [003] diffraction line clearly demonstrates the decrease in *c* lattice parameter.

It is clear that the presently studied flux free $Fe_{1-x}Ni_xSe_{0.5}Te_{0.5}$ single crystals are big in Size (Fig.1), their growth is laminar type (Fig. 2a) and composition (Fig. 2b) is near perfect. Their on surface single crystal XRD showed that they grow in [00l] direction. Detailed powder XRD and Reitveld analysis revealed clear decrease in lattice parameters and volume with increase of Ni content in $Fe_{1-x}Ni_xSe_{0.5}Te_{0.5}$ single crystals series.

The resistivity verses temperature plots ($\rho$-$T$) for $Fe_{1-x}Ni_xSe_{0.5}Te_{0.5}$ single crystals series in temperature range of 2 to 50K are given in Table 5. The normal state resistivity is normalized at 300K; hence forth ideally the plots shown are for $\rho/\rho_{300}$ versus temperature. The pristine i.e., x = 0.0 crystal shows superconducting transition temperature ($T_c$) being defined as resistivity ($\rho$) = 0 at around 12K. The onset of $T_c$, i.e., the temperature below which the superconducting transition starts is at around 14K for the un-doped pristine crystal. With increase of Ni content in $Fe_{1-x}Ni_xSe_{0.5}Te_{0.5}$, the onset of $T_c$ decreases to around 9K, 6K and 4K for x = 0.01, 0.03 and 0.05 crystals. The $T_c(\rho=0)$ values of 4K and 2K are obtained for x = 0.01 and 0.03 crystals and the x=0.05 crystal did not show the same down to 2K. For higher Ni content i.e. x = 0.07, 0.10 and 0.20 crystals neither the onset of $T_c$ nor the $T_c(\rho=0)$ are seen down to 2K. The deleterious effect of Fe site Ni in $FeSe_{0.5}Te_{0.5}$ superconductor appears to be about -8K/at%. This appears to be the same as being found in case of infamous Zn substitution at Cu and Fe sites in high $T_c$ cuprates [26] and Fe pnictides [27] respectively. As far as normal state conduction process is concerned, though the pristine crystal exhibited metallic behavior, Ni doped ones are semi-metallic. Also though the increase in normal state resistivity is seemingly monotonic with increase in Ni content, the same was not true for x = 0.20 crystal, for which though conduction is semi-metallic but absolute resistivity values are smaller than other Ni doped crystals. Overall, it seems the increase in Ni content in $Fe_{1-x}Ni_xSe_{0.5}Te_{0.5}$ suppresses superconductivity fast and quenches the same at just above 3at% doping of Ni at Fe site.

To further ascertain upon the superconducting properties of $Fe_{1-x}Ni_xSe_{0.5}Te_{0.5}$ crystals, the magneto-transport $\rho(T)H$ data for all superconducting samples are shown in Figs. 6(a-d). The magnetic field applied is perpendicular to *ab* plane and up to 14 Tesla. Precisely, one gets interested to know, if the doped Ni could induce pinning and increase the upper critical field of the pristine $FeSe_{0.5}Te_{0.5}$ superconductor. For pristine sample (Fig.6a), though both $T_c$ (onset) and $T_c(\rho=0)$ decrease monotonically with applied field, the relative decrease is faster for the later. The upper critical field value at zero temperature ($H_{c2}$, 0K) being calculated from standard equations is in range of 100 Tesla. This is being already discussed in detail in ref. 23. The $\rho(T)H$ results for $Fe_{0.99}Ni_{0.01}Se_{0.5}Te_{0.5}$ crystal are shown in Fig. 6(b). Clearly the superconducting transition is quite broad even without applying the magnetic field. Precisely though $T_c$ (onset) is at around 9K the $T_c(\rho=0)$ is at around 4K. With application of magnetic field the superconducting transition further broadens and $T_c(\rho=0)$ state is not seen for above 2 Tesla field. At the same time, though $T_c$ (onset) temperature decreases with field, but seen altogether up to 14 Tesla. Numerically, the $T_c$ (onset) of around 9K without magnetic field decreases to around 5K at 14 Tesla. This shows that though superconductivity is suppressed fast with Ni doping at Fe site in $FeSe_{0.5}Te_{0.5}$ superconductor, the intrinsic nature of this class of superconductivity i.e., robustness against magnetic field is yet intact. Fig. 6(c) depicts the $\rho(T)H$ results for $Fe_{0.97}Ni_{0.03}Se_{0.5}Te_{0.5}$ crystal. As discussed before this crystal just attains $T_c(\rho=0)$ state at 2K, which is the lowest temperature we could go down to on our PPMS. With increase in magnetic field though the $T_c(\rho=0)$ could not be seen down to 2K, the $T_c$ (onset) could be seen right up to as high field as 12 Tesla at around 3K. This further confirms the fact that though superconductivity temperature decreases fast with Ni doping in Fe chalcogenide superconductor, the robustness of the phenomenon yet remains. For $Fe_{0.95}Ni_{0.05}Se_{0.5}Te_{0.5}$ crystal only onset of superconductivity is seen and $T_c(\rho=0)$ was not seen down 2K. With applied magnetic field, though the $T_c$ (onset) decreases to lower temperatures, the relative decrease of the same with field is yet very much comparative to the other superconducting crystals i.e. x = 0.0., 0.01 and 0.03. It is clear from Figs. 6(a-d) that though Ni suppresses superconductivity fast (~8K/at%) in $FeSe_{0.5}Te_{0.5}$ superconductor, the robustness of the phenomenon against magnetic field yet remains intact

Summarily, we have grown successfully the single crystal of series $Fe_{1-x}Ni_xSe_{0.5}Te_{0.5}$ (x=0.0, 0.01, 0.03, 0.05, 0.07, 0.10 and 0.20) without added flux for first time. It is found that Fe site Ni substitution suppresses superconductivity quite fast to the tune of ~ 8K/at%. Further it is seen from high field (14 Tesla) magneto-transport measurements that though $T_c$ decreases fast with Ni doping, the robustness of the superconductivity of the compound yet remains intact against magnetic field.

**Acknowledgement**

The authors would like to thank the Director of NPL-CSIR India for his encouragement. This work is financially supported by a DAE-SRC outstanding investigator award scheme on the search for new superconductors.

**FIGURE CAPTIONS**

**Figure 1:** Photograph of $Fe_{1-x}Ni_xSe_{0.5}Te_{0.5}$ (x=0.0, 0.01, 0.03, 0.05 and 0.10) crystals.

**Figure 2:** (a) SEM images and (b) EDX quantitative analysis for $Fe_{1-x}Ni_xSe_{0.5}Te_{0.5}$ single crystal.

**Figure 3:** Single crystal surface XRD for $Fe_{1-x}Ni_xSe_{0.5}Te_{0.5}$ (x=0.0, 0.01, 0.03, 0.05, 0.07, 0.10 and 0.20) at room temperature

**Figure 4:** Powder XRD patterns for crushed powders of $Fe_{1-x}Ni_xSe_{0.5}Te_{0.5}$ (x=0.0, 0.01, 0.03, 0.05, 0.07, 0.10 and 0.20) single crystals at room temperature. *Inset* view is the expanded [003] plane view for the same.

**Figure 5:** Normalised resistivity ($\rho/\rho_{300}$) versus temperature plots for $Fe_{1-x}Ni_xSe_{0.5}Te_{0.5}$ (x=0.0, 0.01, 0.03, 0.05, 0.07, 0.10 and 0.20) single crystals in temperature range of 2 to 50K.

**Figure 6:** (a) Magneto-transport $\rho(T)H$ plots in superconducting state up to 14 Tesla for (a) $FeSe_{0.5}Te_{0.5}$, (b) $Fe_{0.99}Ni_{0.01}Se_{0.5}Te_{0.5}$, (c) $Fe_{0.97}Ni_{0.03}Se_{0.5}Te_{0.5}$ and (d) $Fe_{0.95}Ni_{0.05}Se_{0.5}Te_{0.5}$ single crystals.

**Table 1:** Rietveld refined lattice parameters and coordinate positions for $Fe_{1-x}Ni_xSe_{0.5}Te_{0.5}$ single crystals

|  | x= 0.00 | x=0.01 | x=0.03 | x=0.05 | x=0.10 | x=0.2 |
|---|---|---|---|---|---|---|
| a=b (Å) | 3.801(2) | 3.800(3) | 3.7999(4) | 3.797(3) | 3.795(3) | 3.784(3) |
| c (Å) | 5.998(4) | 5.9959(3) | 5.9958(4) | 5.979(3) | 5.937(2) | 5.895(3) |
| V(Å$^3$) | 86.725(3) | 86.582(2) | 86.575(3) | 86.2264(3) | 85.5304(3) | 84.408(3) |
| Fe | (3/4,1/4,0) | (3/4,1/4,0) | (3/4,1/4,0) | (3/4,1/4,0) | (3/4,1/4,0) | (3/4,1/4,0) |
| Ni |  | (3/4,1/4,0) | (3/4,1/4,0) | (3/4,1/4,0) | (3/4,1/4,0) | (3/4,1/4,0) |
| Se | (1/4,1/4,0.286) | (1/4,1/4,0.283) | (1/4,1/4,0.279) | (1/4,1/4,0.277) | (1/4,1/4,0.261) | (1/4,1/4,0.253) |
| Te | (1/4,1/4,0.286) | (1/4,1/4,0.283) | (1/4,1/4,0.279) | (1/4,1/4,0.277) | (1/4,1/4,0.261) | (1/4,1/4,0.253) |

Fig. 1

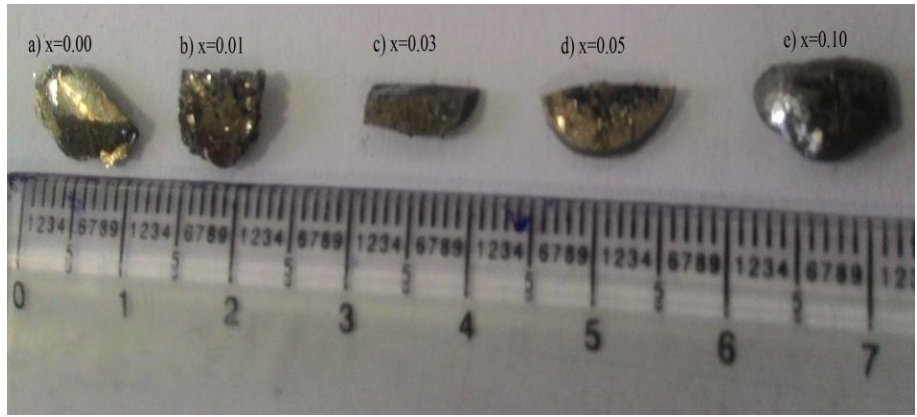

Fig. 2

(a) (b)

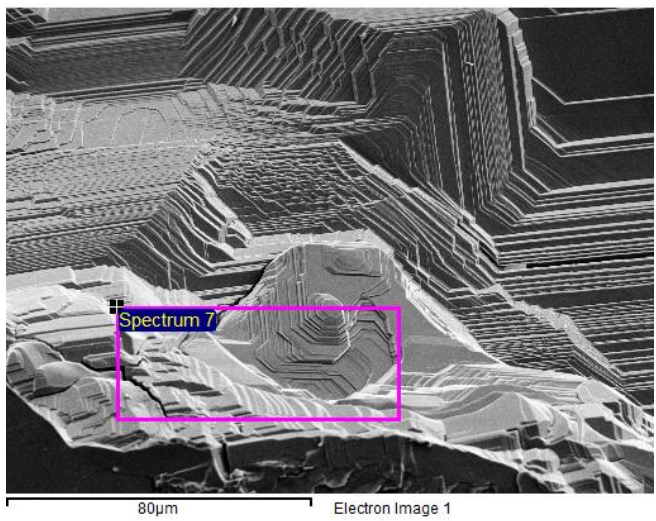
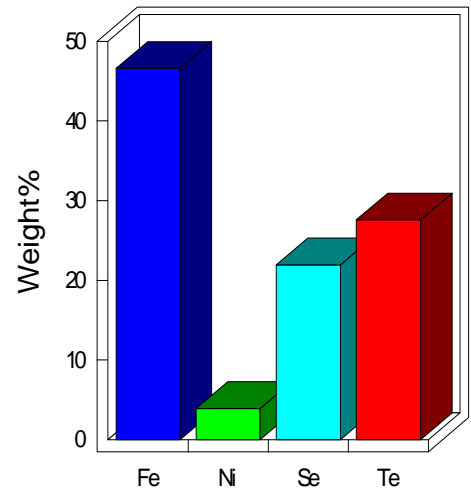

Fig. 3

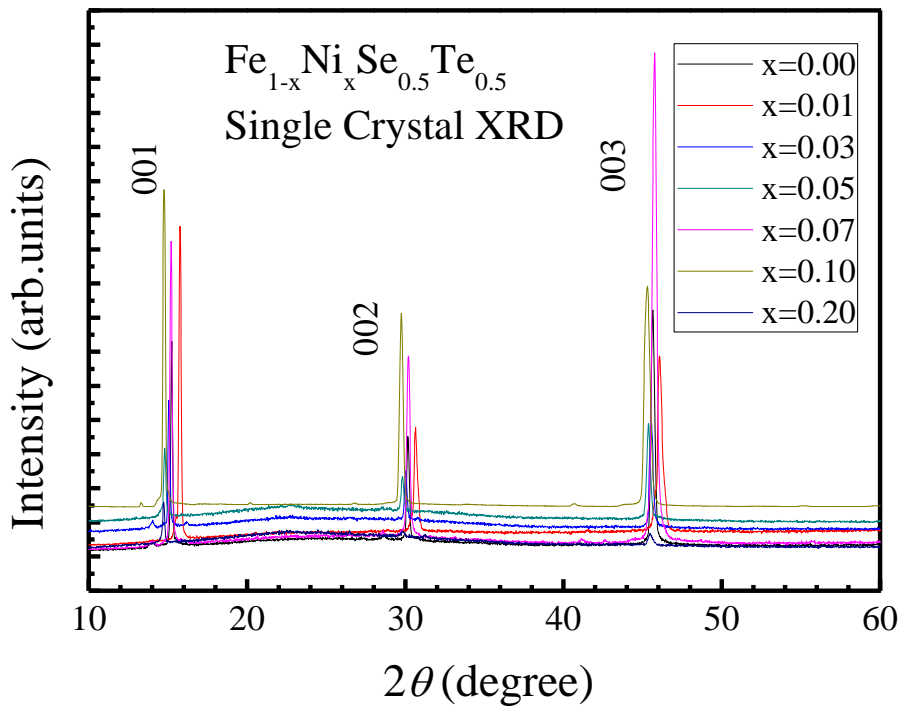

Fig. 4

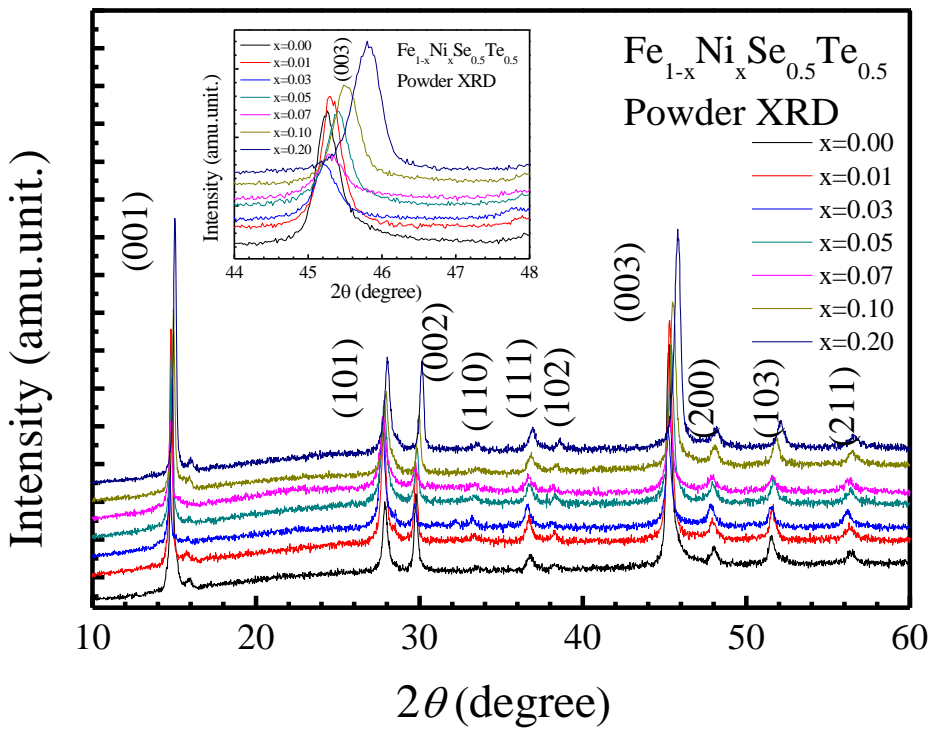

Fig. 5

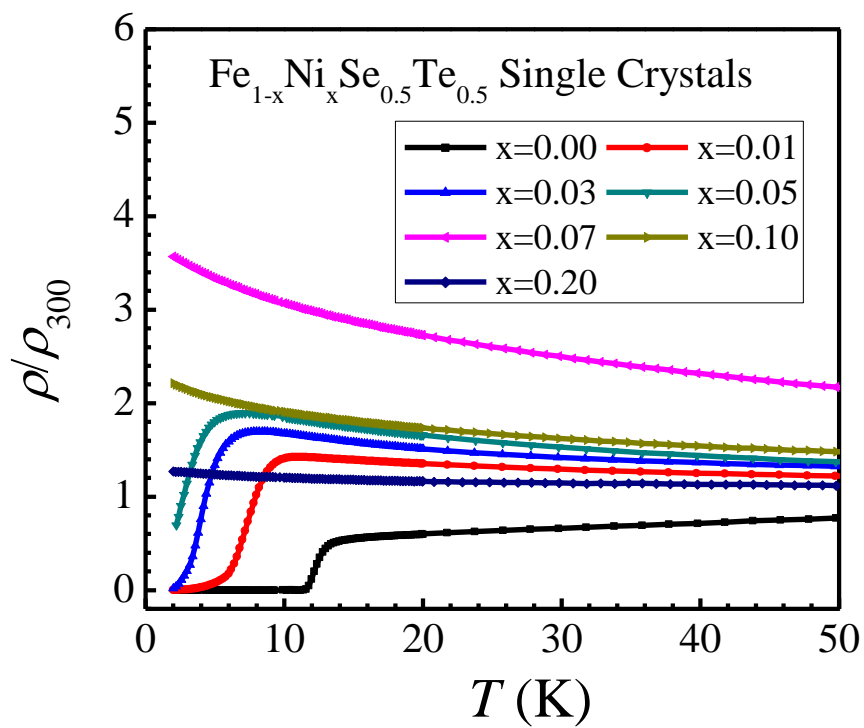

Fig. 6(a)

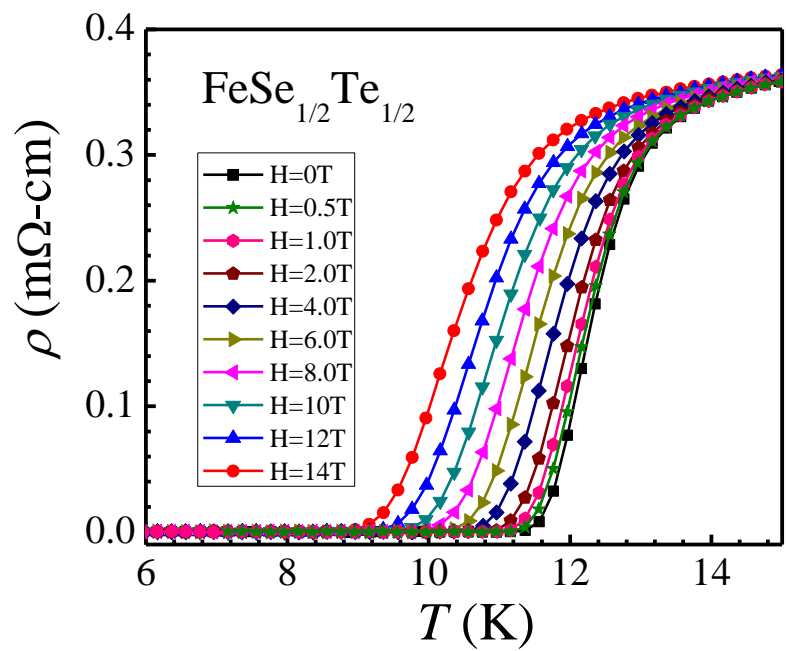

Fig. 6(b)

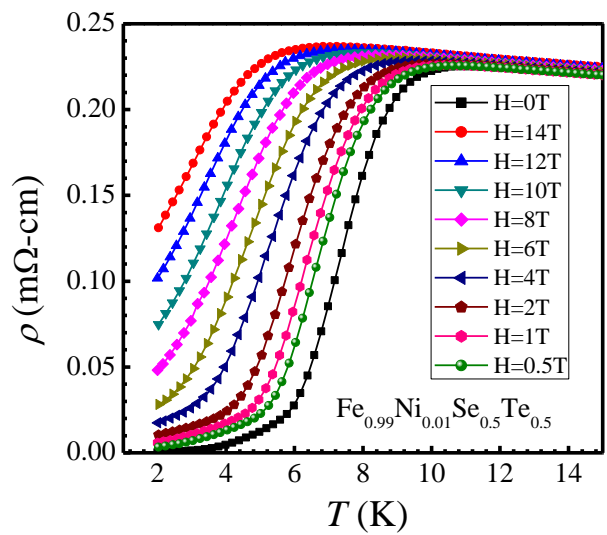

Fig. 6(c)

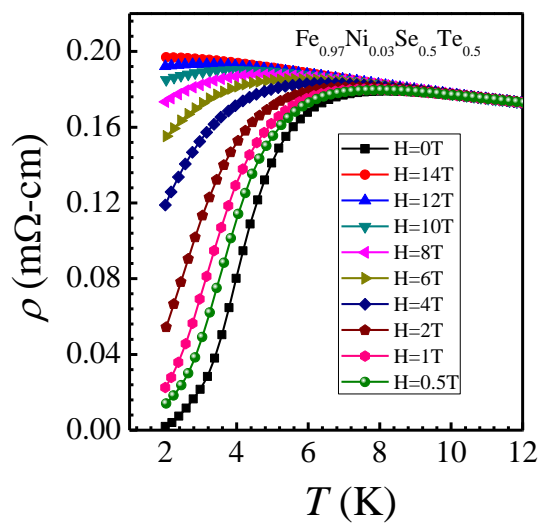

Fig. 6(d)

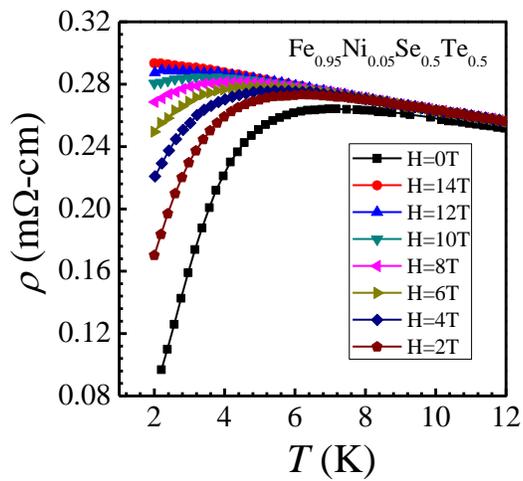

14